\documentstyle[12pt,a4]{article}
\begin{document}
\title{ ON THE DYNAMICS OF RATIONAL SOLUTIONS FOR 1-D GENERALIZED
VOLTERRA SYSTEM}
\author{A. S. C\^arstea\thanks{e-mail: acarst@theor1.ifa.ro}\and{\it Institute
of Atomic Physics, Bucharest, P.O. Box MG-6, Romania}}

\maketitle

\begin{abstract}

The Hirota bilinear formalism and soliton solutions for a  generalized 
Volterra system is presented. Also, starting from the soliton solutions,
we obtain a class of nonsingular rational solutions using the "long wave
limit" procedure of Ablowitz and Satsuma, and appropriate
 "gauge" transformations.
 Their properties are also discussed
and it is shown that these solutions interact elastically with no phase shift.

\end{abstract}

\section{Introduction}

It is known that with the development of soliton theory significant progress
has been made in finding special solutions of integrable nonlinear evolution
equations which include soliton solutions, rational solutions, similarity
solutions and so on, using I. S. T.\cite{ablseg}, Hirota bilinear formalism
\cite{hirsat}, Backlund transformations \cite{colrog} and Wronskian formalism
\cite{grecu}, \cite{nimmo}.

The class of rational solutions was firstly investigated for KdV equation
\cite{airmoser}. Avery simple way to find them was developed by Ablowitz
and Satsuma \cite{ablsat}, and consist in taking the "long wave" limit in the
multisoliton solution. This method was used succesfully to find the rational
solutions also for other completely integrable systems.   
Although, for continuous case, many results have been obtained with respect to
finding them, a relatively less work for discrete systems has been done
\cite{acarst}, \cite{clarkson}.

Unlike the (2+1)-dimensional continuous systems where the dynamics
 and phenomenology of rational solutions ("lump" solutions) are well studied
\cite{freeman}, \cite{konopel}, the dynamics of (1+1)-dimensional nonsingular 
rational solutions was very little investigated \cite{Ono}.
In this paper we'll derive the bilinear formalism for a (1+1)-dimensional
 discrete generalized Volterra system, which represent a coupled system of 
generalized Volterra equations proposed  by M.Wadati
\cite{wadati}, \cite{bullough}. From its soliton solutions we'll obtain a class
of real and complex rational nonsingular solutions using Ablowitz-Satsuma 
limiting procedure. Here the crucial step is the existence of many nontrivial
arbitrary constants and "gauge" transformations. It is shown that
 these solutions collide elastically with no phase shift using a "head-on"
collision of two 1-rational nonsingular solutions. Also their real and complex
structure is preserved. The properties, generalizations and possible
 developments are also discussed.

The paper is organized as follows. In the section 2 is constructed the 
bilinear form of the system by a reduction to a variant of Ablowitz-Ladik
system with nonzero boundary condition. In section 3 the construction of 
rational solutions is presented. In section 4 properties and dynamics are 
discussed and, finally, conclusions are given in section 5. 
  
\section{Bilinear formalism}

The generalized Volterra system under consideration is:
\begin{eqnarray}
\dot Q_{n}=(a+bQ_{n}+cQ_{n}^{2})(R_{n+1}-R_{n-1}) \nonumber\\
\dot R_{n}=(a+bR_{n}+cR_{n}^{2})(Q_{n+1}-Q_{n-1}) 
\end{eqnarray}

with $Q_{n}, R_{n} \rightarrow 0$ as $n$ goes to infinity and $a$, $b$, $c$
are real constants.

For waves propagating in one direction we may put $Q_{n}=R_{n}$ and we'll
find a generalized Volterra equation proposed firstly by M. Wadati
 \cite{wadati}, \cite{bullough}.

By the following transformations:

\begin{eqnarray}
u_{n}=\frac{2c}{\sqrt{4ac-b^{2}}} (Q_{n}+b/2c) \nonumber\\
v_{n}=\frac{2c}{\sqrt{4ac-b^{2}}} (R_{n}+b/2c)\nonumber\\
t \rightarrow t(\frac{4ac-b^{2}}{4c}) 
\end{eqnarray}
and putting $\alpha_{0}=\frac{b}{\sqrt{4ac-b^{2}}}$

the system (1) becomes:
\begin{eqnarray}
\dot u_{n}=(1+u_{n}^{2})(v_{n+1}-v_{n-1}) \nonumber\\
\dot v_{n}=(1+v_{n}^{2})(u_{n+1}-u_{n-1}) 
\end{eqnarray}
with $u_{n}, v_{n} \rightarrow \alpha_{0}$ as $n$ goes to infinity. In order
to have $\alpha_{0}$ real we consider that $c$, $b$ are nonzero
 and $4ac-b^{2}$ is
positive. The system (3) represent a variant of Ablowitz-Ladik system
\cite{ablad}, with nonvanishing boundary conditions.

We are constructing the bilinear form of this system by taking:
\begin{eqnarray}
u_{n}=\alpha_{0}-\frac{i}{2} \frac{d}{dt} \log{(\frac{f_{n}}{g_{n}})}
\nonumber\\
v_{n}=\alpha_{0}-\frac{i}{2} \frac{d}{dt} \log{(\frac{f_{n}'}{g_{n}'})}
\end{eqnarray} 
where we assume that $f_{n}/g_{n}$ and $f_{n}'/g_{n}'$ $\rightarrow const.$ as
$n$ goes to infinity.

Plugging (4) into (3), integrating with respect to $t$ and taking into account
the nonzero boundary conditions we'll get:
$$\tan^{-1}{u_{n}}=\tan^{-1}{\alpha_{0}}-\frac{i}{2}\log{(\frac
{f_{n+1}'g_{n-1}'}{g_{n+1}'f_{n-1}'})} $$
$$\tan^{-1}{v_{n}}=\tan^{-1}{\alpha_{0}}-\frac{i}{2}\log{(\frac
{f_{n+1}g_{n-1}}{g_{n+1}f_{n-1}})}$$
Using (4) again we'll find after some algebra:
\begin{eqnarray}
D_{t}f_{n}g_{n}=2(1+\alpha_{0}^{2})\sinh{D_{n}}f_{n}'g_{n}'\nonumber\\
D_{t}f_{n}'g_{n}'=2(1+\alpha_{0}^{2})\sinh{D_{n}}f_{n}g_{n}\nonumber\\
(\cosh{D_{n}}+i\alpha_{0}\sinh{D_{n}})f_{n}'g_{n}'=f_{n}g_{n}\nonumber\\
(\cosh{D_{n}}+i\alpha_{0}\sinh{D_{n}})f_{n}g_{n}=f_{n}'g_{n}'
\end{eqnarray}
which represent the Hirota form of the system (3). We can see that for 
$\alpha_{0}=0$ we can recover the self-dual network form proposed by Hirota
\cite{hirsat}.

The 1-soliton solution are given by:
\begin{eqnarray}
f_{n}=1+\exp(\eta_{1}+\phi_{1})\nonumber\\
f_{n}'=1+\exp(\eta_{1}+\phi_{1}')\nonumber\\
g_{n}=1+\exp(\eta_{1}+\psi_{1})\nonumber\\
g_{n}'=1+\exp(\eta_{1}+\psi_{1}')
\end{eqnarray}
where
\begin{eqnarray}
\eta_{j}=2\epsilon_{j}(1+\alpha_{0}^{2})\sinh{P_{j}}t+P_{j}n+
\eta_{j}^{0}\nonumber\\
\exp{\phi_{j}}=\frac{\alpha_{0}}{2}K_{j} \frac{\epsilon_{j}+\cosh{P_{j}}}{
\sinh{P_{j}}}+\frac{i}{2}K_{j}\nonumber\\
\exp{\psi_{j}}=\frac{\alpha_{0}}{2}K_{j}\frac{\epsilon_{j}+\cosh{P_{j}}}{
\sinh{P_{j}}}-\frac{i}{2}K_{j}\nonumber\\
\exp{\phi_{j}'}=\frac{\alpha_{0}}{2}K_{j}\frac{1+\epsilon_{j}\cosh{P_{j}}}{
\sinh{P_{j}}}+\frac{i\epsilon_{j}}{2}K_{j}\nonumber\\
\exp{\psi_{j}'}=\frac{\alpha_{0}}{2}K_{j}\frac{1+\epsilon_{j}\cosh{P_{j}}}{
\sinh{P_{j}}}-\frac{i\epsilon_{j}}{2}K_{j}
\end{eqnarray}
where the index $j$ labels the number of the soliton i.e. for 1-soliton
solution $j=1$, for 2-soliton $j=1, 2$ and so on and $\epsilon_{j}=\pm 1
$ indicates the propagation direction.
 On the other hand we've
choosen the phases such that any multisoliton solution to be real. $K_{j}$
represents arbitrary real constants which does not depend on $n$ or $t$. The
procedure of finding rational solutions relies mainly on this arbitrariness.

Now we can proceed to the 2-soliton solution which has the form:
\begin{eqnarray}
f_{n}=1+\exp(\eta_{1}+\phi_{1})+\exp(\eta_{2}+\phi_{2})+
      \exp(\eta_{1}+\eta_{2}+\phi_{1}+\phi_{2}+A_{12})\nonumber\\
f_{n}'=1+\exp(\eta_{1}+\phi_{1}')+\exp(\eta_{2}+\phi_{2}')+
      \exp(\eta_{1}+\eta_{2}+\phi_{1}'+\phi_{2}'+A_{12})\nonumber\\
g_{n}=1+\exp(\eta_{1}+\psi_{1})+\exp(\eta_{2}+\psi_{2})+
      \exp(\eta_{1}+\eta_{2}+\psi_{1}+\psi_{2}+A_{12})\nonumber\\
g_{n}'=1+\exp(\eta_{1}+\psi_{1}')+\exp(\eta_{2}+\psi_{2}')+
       \exp(\eta_{1}+\eta_{2}+\psi_{1}'+\psi_{2}'+A_{12})
\end{eqnarray}
where 
$$\exp{A_{12}}=\left[\frac{\sinh{(\frac{P_{1}-P_{2}}{2})}}
               {\sinh{(\frac{P_{1}+P_{2}}{2})}}\right]^{2}$$
for solitons propagating in the same directions ($\epsilon_{1}\epsilon_{2}=1$)
 and:
$$\exp{A_{12}}=\left[\frac{\cosh{(\frac{P_{1}-P_{2}}{2})}}
               {\cosh{(\frac{P_{1}+P_{2}}{2})}}\right]^{2}$$
for solitons propagating in opposite directions($\epsilon_{1}\epsilon_{2}=-1$)

Using the classical procedure we can construct any N-soliton solution but for
our purposes 1-soliton and 2-soliton solutions are suficient.

\section{Rational Solutions}
The "long wave limit" method of Ablowitz and Satsuma we'll use henceforth 
consists in three main steps. In order to obtain a N-rational solution from
N-soliton solution we have to:
\begin{itemize}
\item expand the Hirota form of the N-soliton solution in power series of
$P_{j}$ with $j=1,...,N$
\item choose the phase constants such that all $O(P_{j}^{k})$ with 
$k$ less than N to be zero.
\item making $ P_{j}\rightarrow 0$ we recover the N-rational solution from the
$O(P_{j}^{l})$ terms with $l \geq N $
\end{itemize}

We are focusing on $f_{n}$ and $g_{n}$. The procedure for $f_{n}'$ and 
$g_{n}'$
works in the same manner. Thus, for 1-rational solutions we'll expand 
$f_{n}$ and $g_{n}$ given by (6). We'll use the following notations:
$$\theta_{j}=2\epsilon_{j}(1+\alpha_{0}^{2})t+n$$
$$\zeta_{j}=\exp{\eta_{j}^{0}}$$
$$P_{j}=\delta \bar p_{j}$$ 
and the expansions will be in power of $\delta$ and $\bar p_{j}$ are $O(1)$
terms.

So, up to $O(\delta)$:
$$f_{n}=1+\zeta_{1}(1+\delta \bar p_{1} \theta_{1}+O(\delta^{2}))
\left[\frac{\alpha_{0}( \epsilon_{1}+1)}{2 \bar p_{1}}K_{1} \delta^{-1}+
\frac{i}{2}K_{1}+\frac{K_{1} \bar p_{1} \alpha_{0}}{12}\delta+
O(\delta^{2})\right]$$
$$g_{n}=(f_{n})^{*}$$
where (*) means changing the sign of the terms multiplicated with $i$ 
(complex conjugate as if $\zeta_{j}$ would be real for $j=1, 2$).

For $\epsilon_{1}=1$ we $must$ take$ K_{1}$ proportional with $\delta$. 
To cancel $O(1)$ terms we're taking $\zeta_{1}=-1/\alpha_{0}$. Thus when 
$\delta \rightarrow 0$ we obtain:

\begin{equation}\label{1ratreal}
u_{n}= \alpha_{0}-\frac{i}{2} \frac{d}{dt} \log{(\frac{\theta_{1}+i/2
\alpha_{0}}
{\theta_{1}-i/2\alpha_{0}})}= \alpha_{0}- \frac{4\alpha_{0}(1+
\alpha_{0}^{2})}
{4\alpha_{0}^{2}\theta_{1}^{2}+1}
\end{equation}
In the same manner $v_{n}=u_{n}$ so, up to a constant and time scaling factor,
$$Q_{n}=R_{n} \sim -\frac{4\alpha_{0}(1+\alpha_{0}^{2})}{4\alpha_{0}^{2}
\theta_{1}^{2}+1}$$
which is a nonsingular real rational solution.

For$ \epsilon_{1}=-1$ the situation is dramatically changed. In this case the 
expansions becomes:
$$f_{n}=1+\frac{i}{2}K_{1}\zeta_{1}+\frac{i}{2}K_{1} \zeta_{1} \bar p_{1}(
\theta_{1}-i\alpha_{0}/2)\delta+O(\delta^{2})$$
$$g_{n}=(f_{n})^{*}$$

We have no divergent terms in $\delta$ so we can take $K_{1} \sim O(1)$. But
in this case the $O(1)$  terms appear to be complex conjugate in $f_{n}$ and 
$g_{n}$. Accordingly there is no $\zeta_{1}$ to cancel both of them. To cope
with this situation, we can see that $u_{n}$ and $v_{n}$ are
 invariant under the following "gauge" transformation:
\begin{equation}\label{gauge}
\frac{f_{n}}{g_{n}} \rightarrow \frac{f_{n}}{g_{n}}\Lambda
\end{equation}
where $\Lambda$ is an arbitrary complex function which does not depend on t.

Now, taking $\zeta_{1}=2i/K_{1}$ we'll cancel the $O(1)$ terms only in $f_{n}$.
So,
$$f_{n}=- \bar p_{1} (\theta_{1}-i\alpha_{0}/2)\delta +O(\delta^{2})$$
$$g_{n}=2+\bar p_{1} (\theta_{1}+i\alpha_{0}/2)\delta +O(\delta^{2})$$
Choosing $\Lambda=1/\delta$ and then $\delta \rightarrow 0$ we'll find:
$$\frac{f_{n}}{g_{n}}=-\frac{1}{2} \bar p_{1} 
                        (\theta_{1}-\frac{i\alpha_{0}}{2})$$  
and:
\begin{eqnarray}
u_{n}=\alpha_{0}-\frac{i}{2}\frac{d}{dt}
      \log{(\theta_{1}-i\alpha_{0}/2)}\nonumber\\
v_{n}=(u_{n})^{*}
\end{eqnarray}  
which represents complex solitary waves. The physical significance of this 
type of solutions is not clear.
The form of $Q_{n}$ and $R_{n}$ will be given up to constants:
$$Q_{n}=(R_{n})^{*} \sim -\frac{\alpha_{0}}{2} \frac{1+\alpha_{0}^{2}}
{\theta_{1}^{2}+\alpha_{0}^{2}/4}+i\frac{\theta_{1}(1+\alpha_{0}^{2})}
{\theta_{1}^{2}+\alpha_{0}^{2}/4}$$ 

In order to study "head-on" collision of the solutions (9) and (11) we'll
construct the 2-rational solution from 2-soliton solution with solitons 
having opposite directions i. e. $f_{n}$ and $g_{n}$ are given by (8)
and phase factors by (7) with $\epsilon_{1} \epsilon_{2}=-1$.
Expanding $f_{n}$ and $g_{n}$ up to $O(\delta^{2})$ we'll take
 $ K_{1} \sim O(\delta)$ in order to cut divergent behaviour of $\delta^{-1}$
type in $\exp{\phi_{1}}$
$$f_{n}=M_{0}+M_{1}\delta+M_{2}\delta^{2}+O(\delta^{3})$$
$$g_{n}=(f_{n})^{*}$$
where:
$$M_{0}=1+\zeta_{1}\alpha_{0}+\frac{i}{2}K_{2}\zeta_{2}+\frac{i}{2}K_{2}
\zeta_{1}\zeta_{2}$$
\begin{eqnarray}
M_{1}=\frac{i}{2}\bar p_{1} \zeta_{1}+\alpha_{0} \bar p_{1} \zeta_{1}
\theta_{1}+\frac{1}{4}\alpha_{0}\zeta_{2}K_{2} \bar p_{2}+\frac{i}{2}
 \bar p_{2} \zeta_{2} \theta_{2}K_{2}+\nonumber\\
+\frac{K_{2}}{4}(\alpha_{0}^{2}\frac{\bar p_{2}}{\bar p_{1}} -1) \bar p_{1}
\zeta_{1}\zeta_{2}+\frac{i}{2}\alpha_{0}K_{2}\zeta_{1}\zeta_{2}
(\bar p_{1} \theta_{1} +\bar p_{2} \theta_{2})
\end{eqnarray}
The expresion of$ M_{2}$ is very complicated. Anyway after choosing the phase
constants it will be strongly simplified.

Thus we can cancel $O(1)$ terms both in $f_{n}$ and $g_{n}$ by taking
 $\zeta_{1}=-1/\alpha_{0}$  but we cannot do this for $O(\delta)$ terms.
Taking $K_{2}\sim O(1)$ and $\zeta_{2}=2i/K_{2}$ we'll do it for $f_{n}$
only. Introducing a "gauge " transformation in the form $\Lambda=1/\delta$
we will recover $f_{n}$ from $M_{2}$ and $g_{n}$ from $(M_{1})^{*}$.
 So, the rational
solution will be:
\begin{eqnarray}
f_{n}=\theta_{1} \theta_{2}-\frac{3}{4}+\frac{i}{2}
(\frac{\theta_{2}}{\alpha_{0}}-\alpha_{0}\theta_{1})\nonumber\\
g_{n}=\theta_{1}-\frac{i}{2\alpha_{0}}
\end{eqnarray}
and
\begin{eqnarray}
f_{n}'=\theta_{1}+\frac{i}{2\alpha_{0}}\nonumber\\
g_{n}'=\theta_{1}\theta_{2}-\frac{3}{4}-\frac{i}{2}
(\frac{\theta_{2}}{\alpha_{0}}-\alpha_{0}\theta_{1})
\end{eqnarray}

\section{Dynamics and Properties}

The solutions (13) and (14) represents a "head-on" collision of two rational
solitary waves one being real and the other complex, having
 same but opposite velocities. To see this we're keeping $\theta_{1}$ fixed
in (13) and make $t \rightarrow \pm \infty$. We'll find that (13) becomes:
$$u_{n} \rightarrow  \alpha_{0}-\frac{i}{2}\frac{d}{dt}
\log{(\frac{\theta_{1}+i/2\alpha_{0}}{\theta_{1}-i/2\alpha_{0}})}$$
which represent exactly the real solitary wave(\ref{1ratreal}).

Also, keeping $\theta_{2}$ being fixed and making $t\rightarrow \pm \infty$
we'll get:
$$u_{n} \rightarrow \alpha_{0}-\frac{i}{2}\frac{d}{dt}
\log{(\theta_{2}-i\alpha_{0}/2)}$$
which means that we have no phase shift at this "head-on" collision so, we
have a (1+1)-dimensional discrete "lump-type" dynamics. Also, the real and 
complex character is preserved. We don't know exactly if does exist real
 nonsingular
rational solutions for both directions.

\section{Conclusions}

To summarize, we've showed that rational nonsingular solutions both real and
complex exists even for discrete systems. So this system posses both solitons
and nonsingular (1+1)-dimensional discrete rational solutions
 decaying to zero as $O(n^{-2})$
and their character being real and complex, depending on the propagation
direction.

We've discussed the dynamics of this rational solutions and we've showed that
we have no phase shift at the interaction, specific for the (2+1)-dimensional
continuous "lumps" solutions. Unfortunately we don't know any condition of 
existence for this type of solutions and why they occur in (1+1)-dimensional
systems. Its continuum limit could be viewed as a coupled pair of mKdV
equations with nonzero boundary conditions. We can see that the
 form of 1-rational solution (the real one) has the same structure as those of
mKdV given by Ono \cite{Ono} and Ablowitz-Satsuma \cite{ablsat}, so
 the continuum does not affect it drastically. Probably a setting up of this 
type of solutions in a I. S. T. framework will clarify much more about their
form, dynamics, stability.
\vskip 3 cm
AKNOWLEDGEMENTS: The author would like to express his sincere thanks to Prof. 
J. Hietarinta
and Prof. M. J. Ablowitz for valuable discussions during the Summer School
"Painleve property, one century later", Cargese, Corsica 1996. I am also
very indebted to Prof. R. Conte for his kindness and hospitality which 
allowed  me to participate at this Summer School.

\end{document}